\documentclass[11pt,letterpaper]{article}
\usepackage[margin=0.78in,top=0.72in,bottom=0.72in]{geometry}
\usepackage{fontspec,amsmath,graphicx,enumitem,titlesec,needspace,hyperref}
\hypersetup{hidelinks,pdfauthor={Zimu Huo and coauthors}}
\titleformat{\section}{\large\bfseries}{}{0pt}{}
\titlespacing*{\section}{0pt}{12pt}{4pt}
\setlist[enumerate]{leftmargin=22pt,itemsep=5pt,parsep=0pt,topsep=2pt}
\begin{document}
\begin{center}
{\fontsize{16}{19}\selectfont\bfseries A closed-loop agentic research environment for MRI development within Siemens IDEA platform\par}
\vspace{7pt}
Zimu Huo\textsuperscript{1,2}, Zachary Shah\textsuperscript{1}, Adam B Kerr\textsuperscript{1}, Kawin Setsompop\textsuperscript{1}\par
\vspace{5pt}
{\fontsize{9}{11}\selectfont
\textsuperscript{1}Department of Electrical Engineering, Stanford University, Stanford, United States of America\par
\textsuperscript{2}Department of Biomedical Physics, Stanford University, Stanford, United States of America\par
}
\vspace{4pt}
{\small Contact: \href{mailto:zimuhuo@stanford.edu}{zimuhuo@stanford.edu}}
\end{center}
\section*{Synopsis}
\textbf{Motivation:} Testing new ideas in MRI can require substantial effort to learn sequence programming and reconstruction tools and to debug implementations before the scientific question can be addressed.

\textbf{Goals:} Reduce the implementation and debugging burden of testing MRI ideas while retaining researcher control of scientific decisions.

\textbf{Approach:} We developed an environment where LLM-agents develop sequences in native Siemens-IDEA-platform, evaluate them through Bloch-simulation and reconstruction, and further iteratively refine them using scanner-feedback.

\textbf{Results:} As test cases, the agentic environment was used to implement and validate GIRF-based gradient-calibration-sequence and reconstruction in \textasciitilde{}4hours, and to develop and optimize a rapid 3-second-multi-channel-calibration-scan in \textasciitilde{}12hours of researcher-guided iteration.

\section*{Impact}
We build a framework to leverage LLM agents to automate sequence implementation and debugging which accelerates the translation of novel MRI concepts into scanner experiments, lowering technical barriers and empowering researchers to test complex sequence designs rapidly.

\section*{Introduction}
Testing new ideas in MRI can require substantial effort to learn the software needed to implement them. Researchers may first need to learn vendor-specific sequence-development tools or frameworks such as Pulseq,\textsuperscript{1} implement reconstruction and diagnose software or scanner errors. Prior work like Agent4MR showed that LLM-agents enable sequence development and reconstruction for EPI tasks\textsuperscript{2}. Building on this work and autonomous research systems,\textsuperscript{3} we developed an environment connecting native Siemens IDEA implementation, MR simulation, reconstruction and scanner feedback. Researchers guide the investigation through physical reasoning and review; agents implement and test the proposed methods.

\section*{Methods}
The environment combines a simulation loop with an outer scanner-feedback loop (Figure 1). Agents access tools for editing code, compiling native IDEA sequences, generating protocols, running the vendor sequence simulator, executing GPU Bloch simulation and reconstructing data. Native development retains the vendor’s SAR, PNS, gradient-limit and hardware checks. Sequence events such as, RF, gradient and ADC, drive Bloch simulation using a brain phantom with tissue $T_1$/$T_2$, $B_0$ and multicoil transmit/receive fields. The simulator models RF excitation, relaxation, spin history and off-resonance during readout. Multicoil signals sampled at ADC times are written as synthetic Siemens TWIX raw data, so simulation and scanning share the same reconstruction pipelines.

In this work, we demonstrate the capability of our Agentic environment to rapidly develop sequence and reconstruction for: (1) gradient system characterization using gradient impulse response functions (GIRFs)\textsuperscript{4}, and (2) Rapid field mapping for multichannel Tx array. The first task challenged the agent to help rapidly replicate established data acquisition and reconstruction strategy, whereas the second explored a novel, researcher-guided concept. Here, the fast field mapping optimization addressed tradeoffs among spin history bias, spatial sampling, reconstruction, and scan time, subject to SAR, peak RF voltage, PNS, and gradient limits. Throughout both investigations, researchers supplied physical strategies, reviewed alternatives, and selected designs, while agents implemented and tested the variations between review cycles.

To ensure robust and efficient implementation, agents organized sequence code, Siemens documentation, and pulse-sequence references into linked Markdown notes to build a reusable domain knowledge base (Figure 2). They logged methodologies, implementation details, corrected assumptions, and executable unit tests, linking them to supporting evidence. Updating this knowledge base and its associated tools establishes a mechanism for autonomous self-improvement across iterations.

All the experiments in this work were conducted on the Investigational 7T MAGNETOM Terra.X Impulse Edition scanner (Siemens Healthineers, Erlangen, Germany) with $G_{\max}$=200mT/m, and $S_{\max}$=900T/m/s.

Claude and Codex were used under Stanford’s enterprise agreements, with research data, conversations, and usage logs subject to applicable data-protection requirements and data will not be used for training their models.

\section*{Test example 1: system GIRF characterization}
Here the agent was asked to implement data acquisition and reconstruction algorithms for i) GIRF calibration based on exisiting literature\textsuperscript{4} and ii) a validation spiral scan to assess improvement in reconstruction quality with gradient imperfection information incorporated. In its development and design, the agent was given a realistic synthetic GIRF response, phantom and MRI setup, and asked to design acquisition that would provide good recovery of this GIRF response. Development took approximately 4 hours, including researcher interaction, agent execution, debugging and scanner testing with more details shown in Fig3.

\section*{Test example 2: transmit field mapping}
The agent was tasked to improve the speed and robustness of an existing multichannel parallel transmit $B_1^+$ mapping approach currently available on our 7 T system, which consists of Bloch–Siegert-based $B_1$ mapping of the main circularly polarized (CP) transmit mode, followed by fast low-flip-angle relative $B_1$ mapping of the remaining transmit modes. The current approach uses a 3D-GRE acquisition taking 2.5 minutes, while suffering from $T_1$ bias in long-$T_1$ CSF regions. We tasked the agent to improve mapping speed by transitioning to spiral imaging and to eliminate $T_1$ bias by moving to a 2D multislice acquisition, optimizing parameters across TR, encoding order, slice interleaving, RF-spoiling increment, resolution, and flip angle. More details are shown in Fig4,5. Development took approximately 12 hours, involving multiple successive cycles of agent-driven sequence code generation, simulation-based debugging, and physical validation using both phantom and invivo experiments.

\section*{Discussion and conclusion}
These cases demonstrate a workflow in which researchers direct the scientific investigation while agents carry out the implementation and debugging. The author also explored fully autonomous research, in which agents independently review the literature, conduct analyses, and generate new research ideas. Although these experiments yielded no substantive findings, more capable future models may make this approach more productive.

\needspace{8\baselineskip}
\section*{Acknowledgements}
This work was supported in part by NIH research grants: R01 MH116173; R21 EB038677; R01HD114719; R01 EB033206; U24NS129893; NIH Shared Instrumentation Grant S10-OD036337 and the C-ShaRP Instrumentation \& Enhancement Program at Stanford University.

Claude and Codex were used to assist with coding.

\section*{References}
{\fontsize{9}{11}\selectfont
\begin{enumerate}
\item Layton KJ, Kroboth S, Jia F, et al. Pulseq: A rapid and hardware-independent pulse sequence prototyping framework. Magn Reson Med. 2017;77(4):1544–1552. doi:10.1002/mrm.26235.
\item Zaiss M, Aly A, Endres J, et al. Agentic MR sequence development: leveraging LLMs with MR skills for automatic physics-informed sequence development. arXiv:2604.13282. 2026. doi:10.48550/arXiv.2604.13282
\item Karpathy A. autoresearch: AI agents running research on single-GPU nanochat training automatically. 2026. \url{https://github.com/karpathy/autoresearch.}
\item Addy NO, Wu HH, Nishimura DG. Simple method for MR gradient system characterization and k-space trajectory estimation. Magn Reson Med. 2012;68(1):120–129. doi:10.1002/mrm.23217.
\end{enumerate}
}
\clearpage
\section*{Figures}
\vspace*{8pt}
\begin{center}\includegraphics[width=\linewidth,height=0.70\textheight,keepaspectratio]{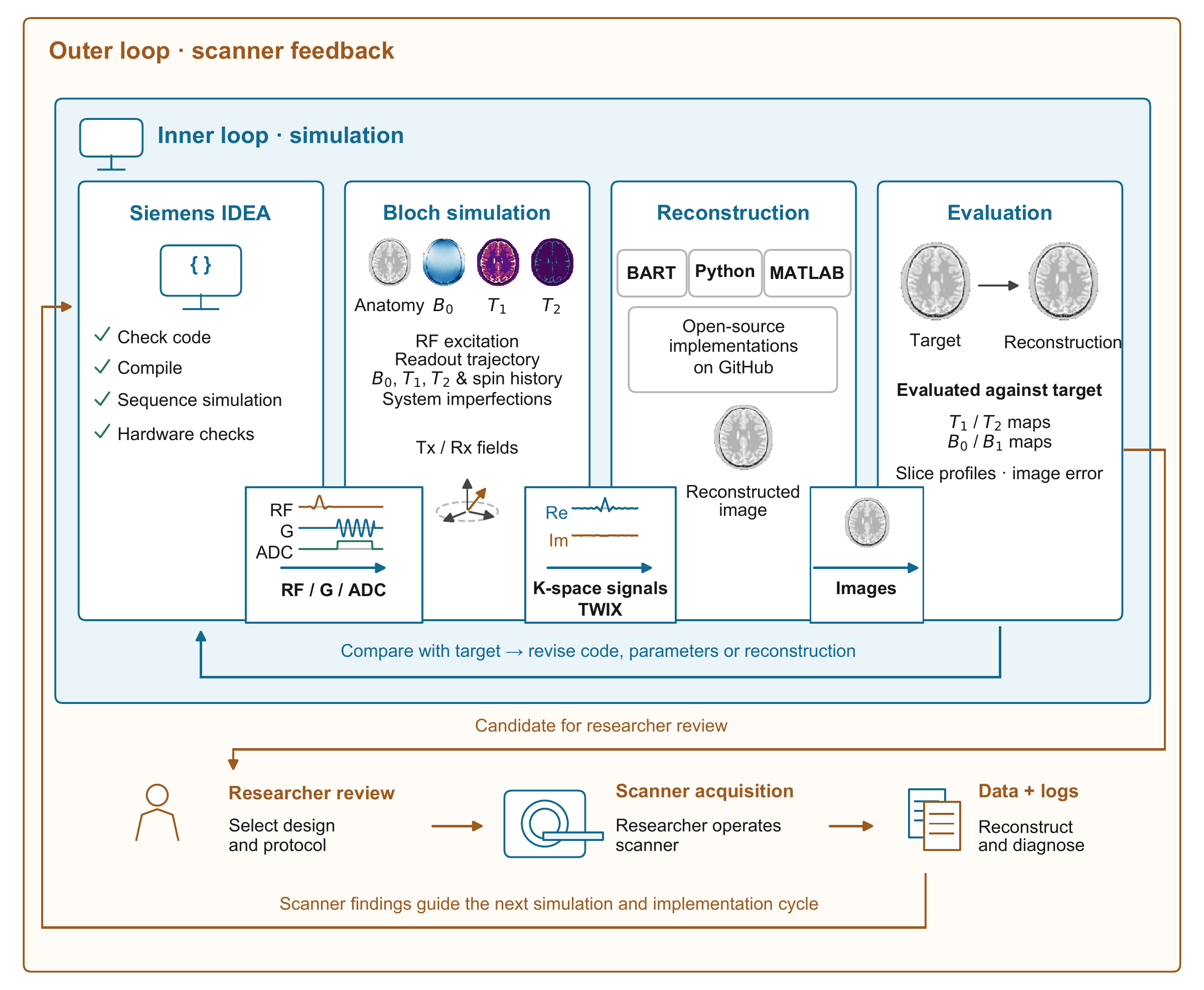}\end{center}
\vspace{8pt}
{\fontsize{10.5}{13}\selectfont\textbf{Figure 1.} Closed-loop agentic workflow combining digital simulation (inner loop) and scanner feedback (outer loop). Agents edit sequence and reconstruction code, run vendor simulation, drive a GPU Bloch simulator to generate synthetic TWIX files, and compare results against ground truth. Researchers review candidates, acquire scanner data, and supply logs to guide agent diagnosis and repair.\par}
\clearpage
\vspace*{8pt}
\begin{center}\includegraphics[width=\linewidth,height=0.70\textheight,keepaspectratio]{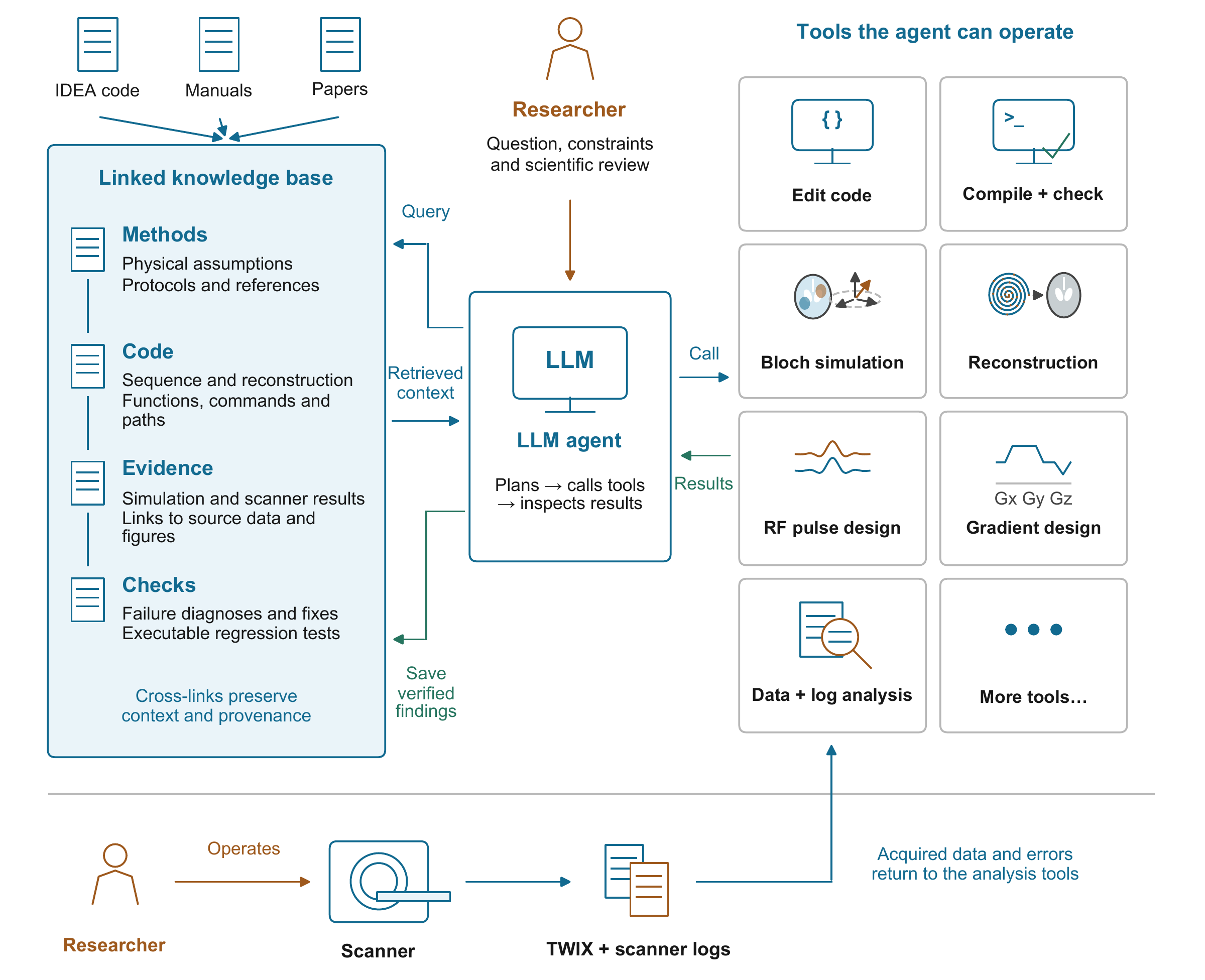}\end{center}
\vspace{8pt}
{\fontsize{10.5}{13}\selectfont\textbf{Figure 2.} Knowledge base construction and iteration. Linked Markdown notes integrate sequence code, documentation, and methods. Agents consult these notes during implementation and record verified findings, corrected assumptions, and regression checks to ensure continuous self-improvement across design cycles.\par}
\clearpage
\vspace*{8pt}
\begin{center}\includegraphics[width=\linewidth,height=0.70\textheight,keepaspectratio]{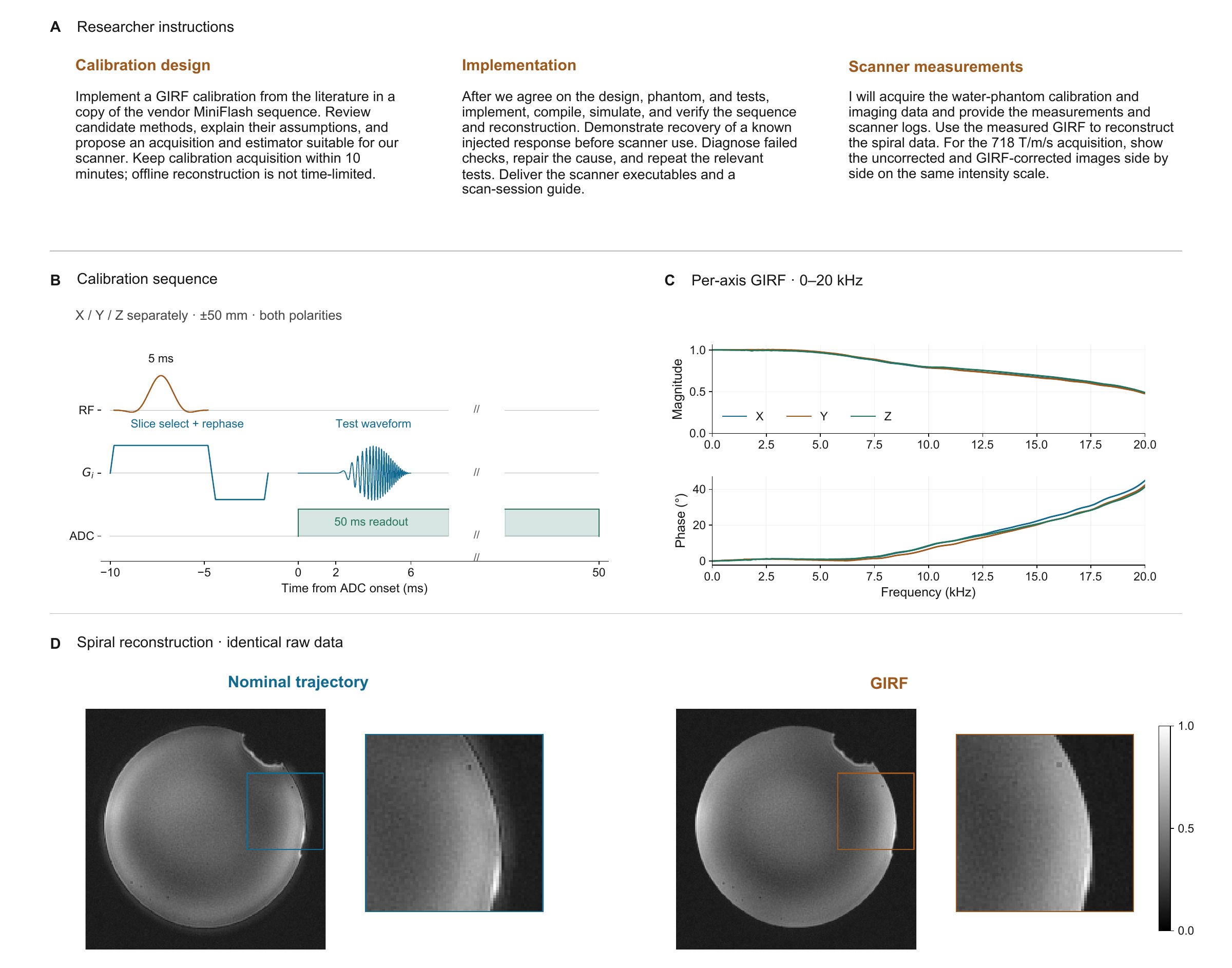}\end{center}
\vspace{8pt}
{\fontsize{10.5}{13}\selectfont\textbf{Figure 3.} GIRF calibration and spiral reconstruction results. Spiral sequence has 1-mm slice thickness, 1 x 1mm resolution over a 256 x 256 mm FOV (Gradient limits: 100 mT/m, 718 T/m/s).\par}
\clearpage
\vspace*{8pt}
\begin{center}\includegraphics[width=\linewidth,height=0.70\textheight,keepaspectratio]{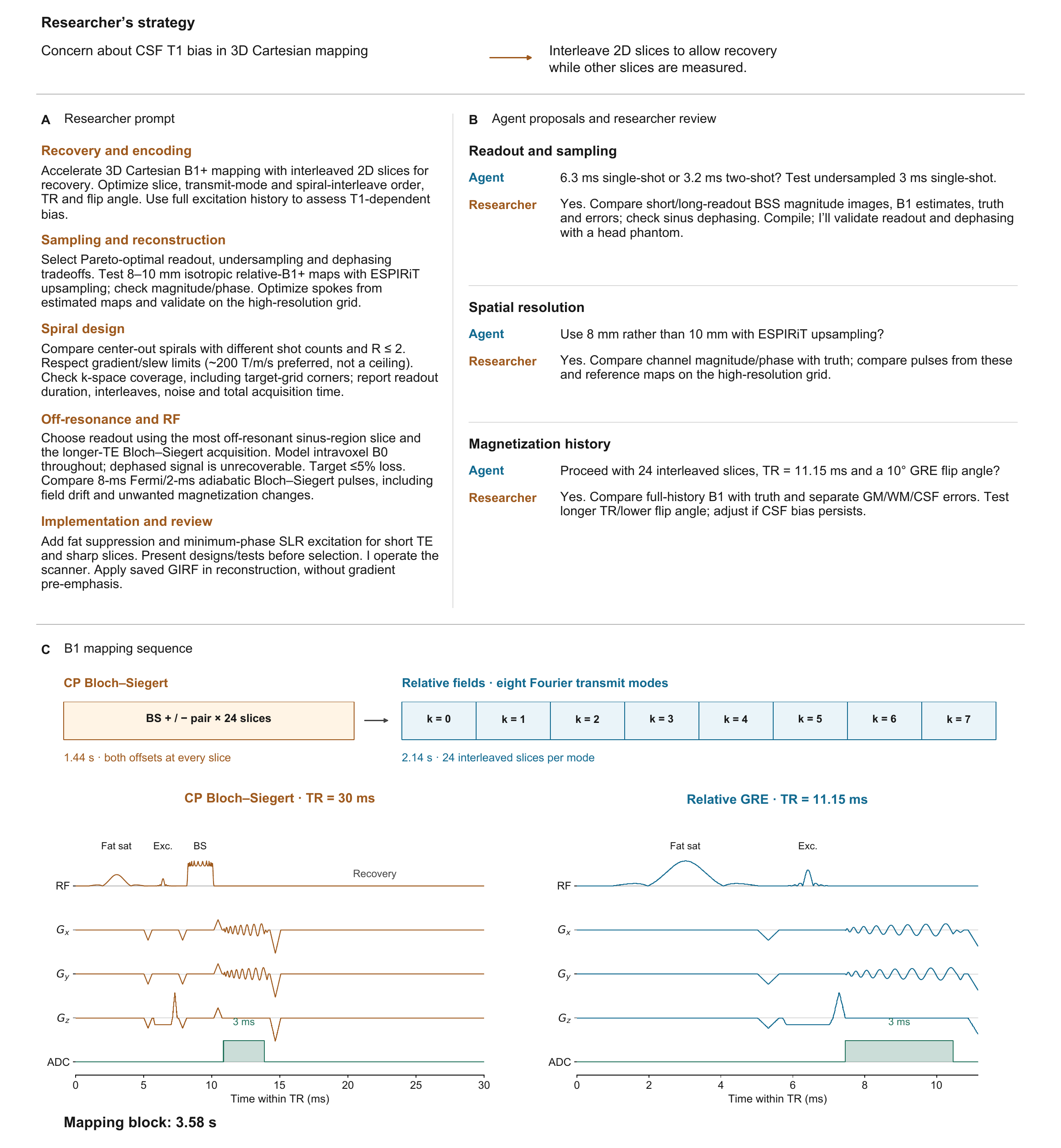}\end{center}
\vspace{8pt}
{\fontsize{10.5}{13}\selectfont\textbf{Figure 4.} Researcher–agent collaborative iteration in $B_1^+$ acquisition and reconstruction design.\par}
\clearpage
\vspace*{8pt}
\begin{center}\includegraphics[width=\linewidth,height=0.70\textheight,keepaspectratio]{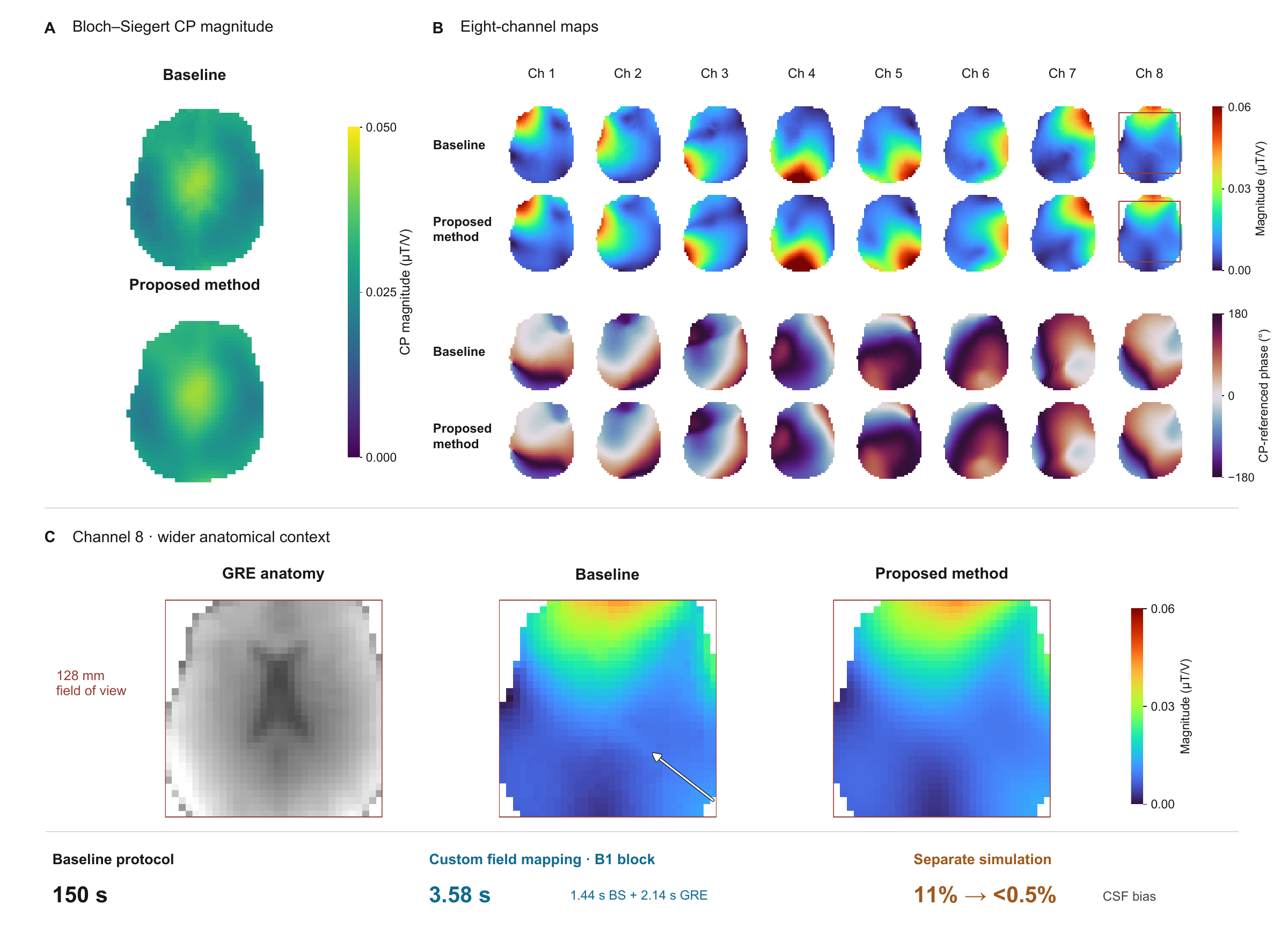}\end{center}
\vspace{8pt}
{\fontsize{10.5}{13}\selectfont\textbf{Figure 5.} Baseline versus proposed measurement of relative magnitude and phase across eight transmit channels. The red arrow marks the CSF region. In separate simulations, CSF bias decreased from 11\% with the baseline protocol to below 0.5\% with the custom interleaved multislice field-mapping sequence co-designed by the researcher and agent.\par}
\end{document}